\magnification\magstep2
  
  \parskip = 0.4 true cm
  \pageno = 1
  \baselineskip = 0.85 true cm
  \vsize = 22 true cm
  \hsize = 16 true cm
  \def\sa{\vskip 0.30 true cm}
  \def\sb{\vskip 0.60 true cm}

  \def\avant{\vskip 1.05   true cm}
  \def\apres{\vskip 0.15   true cm}

    \null \hfill {\bf LYCEN 9341} \break 
    \null \hfill {\bf July 1993}

   \sb
   \sb
   \apres 

\centerline {{\bf Multi-Photon Laser Spectroscopy of Transition 
Ions in Crystals:}}

\centerline {{\bf Recent Progress in the Use of Symmetry Considerations}$^*$}

\sb
\sa

\centerline {\bf M.~Kibler} 

\sa

\centerline {{\it Institut de Physique Nucl\'eaire,}}
 
\centerline {{\it IN2P3-CNRS et Universit\'e Lyon-1,}}

\centerline {{\it 43 Boulevard du 11 Novembre 1918,}} 

\centerline {{\it F-69622 Villeurbanne Cedex, France}}

\sb
\sb
\sb

\noindent {\bf Abstract}~-~The role of symmetry adaptation 
techniques in multi-photon spectroscopy of partly-filled 
shell ions in crystals is briefly reviewed. This leads to 
an intensity formula which is discussed from a qualitative 
point of view.  

\vskip 1.8 true cm 

\noindent {$^*$ Talk given at the ``International 
                  Workshop on Laser Physics (LPHYS-93)'', 
Volga Laser Tour~'93, Russia, 27 June~-~3 July 1993. 
Published in {\bf Laser Physics 4 (1994) 38}.} 

\vfill\eject 
\baselineskip = 0.95 true cm

\centerline {1. INTRODUCTION} 

\apres

In recent years, 
electronic spectroscopy of transition ions (transition-metal 
ions and rare-earth ions) in crystalline environments 
has been the object of important developments. 
In particular, the advent of tunable dye lasers 
has made possible to perform two-photon spectroscopy experiments.   
(In $p$-photon spectroscopy, $p$ photons {\it simultaneously} occur 
in the light-matter interaction. For example, $p$-photon 
absorption spectroscopy corresponds to the {\it simultaneous 
absorption} of $p$ photons between an initial and a final level 
and has to be distinguished from the {\it sequential absorption} 
of $p$ photons between the two levels.) Two-photon spectroscopy 
($p=2$) with laser beams 
is now a technique complementary of one-photon spectroscopy 
($p=1$). It 
gives many informations on the structural (and dynamical) 
properties of ions in crystals. More specifically, two-photon 
spectroscopy permits to reach excited levels which cannot be 
reached with one-photon spectroscopy. In addition, by playing 
with the polarization of the two involved photons, one may obtain 
interesting selection rules. Similar interests exist in 
principle for $p$-photon spectroscopy 
with $p > 2$ although experiments are hardly 
feasible (except for ions in vapors) when $p \ge 4$. 

Since the pioneer works by Kaiser and Garrett [1] and Axe [2] 
in the sixties, there have been many progress, 
from an experimental and a theoretical viewpoint, 
in two-photon 
spectroscopy of $d^N$ and $f^N$ ions in crystals. 
From the theoretical side, the progress 
have been achieved, 
for both parity-allowed and parity-forbidden transitions, 
in two directions: 
 (i) the elaboration of more and more sophisticated models and 
(ii) the systematic use of symmetry considerations [3~-~20] 
based on symmetry adaptation techniques [21,~22]. 
In the present paper, we shall be concerned mainly with point (ii). 
More specifically, we shall give a brief review on the use 
of symmetry considerations, 
in conjunction with sophisticated models, 
for the description of 
electric-dipole two-photon transitions. 
A more detailed exposure may be found in 
[9,~10,~12,~13] for parity-forbidden transitions and in 
      [11~-~13] for parity-allowed   transitions (see also 
Ref.~[23]). 

For the sake of 
generality, we shall give the relevant intensity formulas for 
$p$-photon (electric-dipole) transitions. The case $p=2$ shall 
follow as a particular case. 

\avant

\centerline {2. INTENSITY FORMULA}

\apres

The electric-dipole moment is an odd operator. Therefore,  
for $p$ arbitrary, parity-allowed   transitions correspond to 
either $n \ell^N \to n \ell^N$ transitions for $p$ even 
or     $n \ell^N \to n \ell^{N-1} n' \ell'$ transitions with 
                        $\ell + \ell'$ odd for $p$ odd 
             while parity-forbidden transitions correspond to 
either $n \ell^N \to n \ell^N$ transitions for $p$ odd 
or     $n \ell^N \to n \ell^{N-1} n' \ell'$ transitions with 
                        $\ell + \ell'$ odd for $p$ even. 

Let us consider a $p$-photon (absorption) transition between an 
initial level $i$ of symmetry $\Gamma $ and a 
  final level $f$ of symmetry $\Gamma'$. 
(Here, $\Gamma$ and $\Gamma'$ stand for two 
irreducible representations 
of        the group $G  $ for $N$ even or 
of its spinor group $G^*$ for $N$  odd. We note $G$ the point 
symmetry group of the ${\ell}^N$ ion in its environment. 
Furthermore, we use $\gamma $ and 
                    $\gamma'$ to label the Stark 
components of the levels of symmetry $\Gamma $ 
                                 and $\Gamma'$, respectively.) 
The transition matrix element 
$M_{i(\Gamma \gamma) \to f (\Gamma' \gamma')}$ can be 
calculated in the framework of the following approximations:

(i) we restrict ourselves to the electric-dipole 
(i.e., great wavelength) approximation,  

(ii) single mode excitations (with given energy, 
wave number and polarization vector for each of the $p$ 
photons) are used for describing the radiation field, 

(iii) only $p$-th order mechanisms arising from time-dependent 
perturbation theory are considered, 

(iv) a quasi-closure approximation (which amounts to ignore 
the internal structure of the intermediate configurations) is 
used, 

(v) the only good quantum numbers are 
$\Gamma \gamma$ and $\Gamma' \gamma'$ 
(and the parity for parity-allowed transitions) 
for the initial and final 
state-vectors, respectively, 

(vi) if necessary, as for parity-forbidden transitions, 
$q$-th order mechanisms ($q \ge 1$) arising from 
time-independent perturbation theory are considered, 

(vii) we use a weak-field basis adapted to the chain 
$SU(2) \supset G^*$ for describing the state-vectors and 
interactions.

From recoupling techniques, it can be 
shown that (under the above-mentioned hypotheses) the 
transition matrix element 
$M_{i(\Gamma \gamma) \to f (\Gamma' \gamma')}$ 
is given by
$$
M_{ i(\Gamma \gamma) \to f(\Gamma' \gamma') } 
= ( f \Gamma' \gamma' \vert H_{\rm eff} \vert i \Gamma \gamma ), 
\eqno (1)
$$
where $H_{\rm eff}$ is a model-dependent effective operator. 
The most general form of $H_{\rm eff}$ turns out to be
$$
\eqalign{
H_{\rm eff} = 
\sum_{k_1 k_2 \cdots k_{p-1}} \, \sum_{t} \, \sum_{k_S k_L} \, \sum_{k} \, 
&   C[k_1 k_2 \cdots k_{p-1} ; t ; k_S k_L ; k ] \cr 
& \left( \{ {\bf P}^{(k_{p-1})} 
            {\bf X}^{(t      )} \}^{(k)} . \, 
            {\bf W}^{(k_Sk_L)k} \right),
}
\eqno (2)
$$
where $( \, . \, )$ is a scalar product involving electronic 
(${\bf W}^{(k_Sk_L)k}$) 
and nonelectronic (${\bf P}^{(k_{p-1})}$ and ${\bf X}^{(t)}$) tensors. 
In Eq.~(2), 
${\bf P}^{(k_{p-1})}$ is the polarization tensor 
$$
 {\bf P}^{(k_{p-1})} \> = \> \{ \cdots \{ \{     {\bf E}_1 
                                                 {\bf E}_2 
                           \}^{(k_1    )}        {\bf E}_3
                           \}^{(k_2    )} \cdots {\bf E}_p 
                           \}^{(k_{p-1})} 
\eqno (3)
$$
that describes the coupling of the unit polarization 
vectors ${\bf E}_i$ ($i=1, 2, \cdots, p$). 
In addition, ${\bf W}^{(k_Sk_L)k}$ and 
             ${\bf X}^{(t      )}$ are 
tensors relative to the ion and its environment: 
${\bf W}^{(k_Sk_L)k}$ is an electronic tensor (of 
spin    degree $k_S$, 
orbital degree $k_L$ and 
total   degree $k$) whereas 
${\bf X}^{(t      )}$ is 
the ligand polarization tensor or the crystal-field 
tensor (of the degree $t$). Finally, the expansion coefficients 
$C$ in (2) are calculable in an {\it ab initio} way.

The matrix element (1) can be easily calculated by using 
Wigner-Racah calculus (i.e., irreducible tensor methods) for a 
chain of groups $SU(2) \supset G^*$ as developed in 
Refs.~[21,~22]. 
The quantity of importance for a comparison 
between theory and experiment is the intensity
strength
$$
S_{ i(\Gamma) \to f(\Gamma') } \; = \; \sum_{\gamma \gamma'} \; 
\left\vert M_{i(\Gamma \gamma) \to f(\Gamma' \gamma')} \right\vert ^2 
\eqno (4)
$$
for the $i(\Gamma) \to f(\Gamma')$ transition. The 
sum over $\gamma$ and $\gamma'$ in Eq.~(4) can be 
effectuated by means of the Racah lemma for the Clebsch-Gordan 
coefficients of the group $SU(2)$ in a $SU(2) \supset G^*$ basis. As a 
final result, we obtain the intensity formula [10~-~13] 
$$
\eqalign{
S_{i(\Gamma) \to f(\Gamma')} \; = \; 
                         \sum_{ \left\{ k   _j \right\} } 
                         \sum_{ \left\{ \ell_j \right\} } 
                         \sum_{r} \sum_{s} 
                       & \sum_{\Gamma''} \cr 
I[ \left\{ k   _j \right\} 
   \left\{ \ell_j \right\}
r s \Gamma'' ; \Gamma \Gamma'] \; 
                       & \sum_{\gamma''} \; 
{P}^{(k   _{p-1})}     _{r        \Gamma'' \gamma''} \; \left( 
{P}^{(\ell_{p-1})}     _{s        \Gamma'' \gamma''}    \right)^* 
}
\eqno (5)
$$
with $1 \le j \le p-1$. The polarization dependence in (5) is 
contained in the two $P$ factors. This dependence, which may 
involve linear and/or circular polarization types, is entirely 
under the control of the experimentalist. The intensity 
parameters $I$ in (5) are given by expressions that depend on: 
(i)   the nature (energy) of the $p$ absorbed photons, 
(ii)  the involved 
configurations via energy factors, wavefunctions and radial 
integrals, 
(iii) recoupling coefficients for 
the group $SU(2)$ and reduced matrix elements,
(iv)  the group $G$ via coupling coefficients and isoscalar 
factors for the chain $SU(2) \supset G^*$, and 
(v) the order $q$ of the (time-independent) mechanism used for 
treating some of the interactions (especially the spin-orbit 
interaction and the crystal-field interaction).  
 
 It is remarkable that the structural form of (5) holds for 
both parity-allowed and parity-forbidden transitions. This 
forms (actually based on group theory) also 
holds when some of the absorbed photons are replaced by 
scattered photons (Raman or Rayleigh scattering). 
 The general form of (5) is also valid for other 
multi-photon processes, as for example the simultaneous 
absorption of several photons, certain by 
electric-dipole absorption and others by 
magnetic-dipole and/or 
electric-quadrupole absorption. Note also that vibronic degrees 
of freedom may be incorporated in (5) (see Ref.~[14]). 

\avant

\centerline {3. CLOSING REMARKS} 

\apres

The intensity parameters $I$ may be calculated from first 
principles. This leads to a very much involved quantum 
chemistry problem. 
Alternatively, they may be considered, at least in 
a first approach, as phenomenological parameters to be adjusted 
on experimental data. It may be also interesting to combine 
the {\it ab initio} and {\it phenomenological} approaches. In 
all approaches the number of $I$ parameters is limited by a set 
of properties and selection rules [10~-~13].
 
Once the number of independent parameters $I$ in the intensity
formula (5) has been determined, we can obtain the polarization dependence 
of the intensity strength 
$S_{i(\Gamma) \to f(\Gamma')}$ by calculating the tensor products 
$P ^{(K)} _{a'' \Gamma'' \gamma''}$ 
(with $K = k_{p-1}, \ell_{p-1}$ and $a'' = r, s$) occurring 
in Eq.~(5). For this purpose, we use the development 
$$
P ^{(K)} _{a''\Gamma'' \gamma''} 
\; = \; \sum^K_{Q = - K} \; 
P ^{(K)} _Q \; 
(K Q \vert K a'' \Gamma'' \gamma'')
\eqno (6)
$$
in terms of the spherical components 
$P ^{(K)} _Q$, 
the coefficients in the development 
(6) being reduction coefficients 
for the chain $O(3) \supset G$. Then, in order to calculate 
$P^{(K)}_Q$, we use developments of the type 
$$
\eqalign{
\left\{ {\bf E}_1 
        {\bf E}_2 \right\} ^{(k_1)} _{q_1} \; = \; 
& (-1)^{k_1 - q_1} \; \sqrt{2 k_1 + 1} \cr 
& \sum^1_{x = - 1} \; 
  \sum^1_{y = - 1} \; 
\pmatrix{
1 &  k_1 & 1\cr\cr
x & -q_1 & y\cr
}
\; (E_1)^{(1)}_x 
\; (E_2)^{(1)}_y 
} 
\eqno (7)
$$
in terms of the spherical components $(E_i)^{(1)}_q$ 
characterizing the circular or linear polarization of the 
$i$-th photon ($i=1,2$). In general, we may have $k_1=0,1,2$. 
However, if the two involved photons are identical (one-color 
beam), $k_1$ assumes only the values 0 and 2. Equation (7) describes 
the polarization tensor for $p=2$. If $p > 2$, repeated couplings 
of the type of the one entering Eq.~(7) yield an expression for 
the polarization tensor. 

In the case where $p=2$, the symmetry adaptation techniques and 
models sketched through Eqs.~(1~-~7) have been successfully 
applied to $f^N$ ions [15~-~18] and $d^N$ ions [19,~20] in crystals. 
As a brief r\'esum\'e, we note the following tendencies. For 
intraconfigurational transitions within the configurations 
$3d^N$ ($N \ne 5$)
and $4f^N$ ($N \ne 7$), second-order mechanisms ($p=2, q=0$) 
are generally sufficient for ``nonhypersensitive'' two-photon 
transitions. For ``hypersensitive'' two-photon transitions (like 
$^5D_0$ $\to$ 
$^7F_0$ transitions within the $4f^6$ configuration in tetragonal 
symmetry), third-order mechanisms ($p=2, q=1$) are required for 
a satisfactory description of the polarization dependence. 
For 
inter\-configurational two-photon 
transitions (like $4f^6 \to 4f^55d$ transitions in tetragonal 
symmetry), it is also necessary to consider at least 
third-order mechanisms ($p=2, q=1$) for a good agreement 
between theory and experiment. 

\avant

\centerline {ACKNOWLEDGMENT}

\apres

The author wishes to thank the Organizing Committee of the 
International Workshop on Laser Physics (LPHYS-93) for inviting 
him to give a talk on the present work at this workhop. 

\vfill\eject

\centerline {REFERENCES}

\apres 
\baselineskip 0.57 true cm

\item{1.} Kaiser, W. and Garrett, C.G.B., 1961, {\it Phys. Rev. 
Lett.}, {\bf 7}, 229.
 
\item{2.} Axe Jr., J.D., 1964, {\it Phys.~Rev.}, {\bf 136}, A42. 

\item{3.} Inoue, M. and Toyozawa, Y., 1965, {\it J. Phys. Soc. 
Japan}, {\bf 20}, 363. 

\item{4.} Bader, T.R. and Gold, A., 1968, {\it Phys.~Rev.}, {\bf 171}, 
997. 

  \item{5.} 
  Apanasevich, P.A., Gintoft, R.I., Korol'kov, V.S., 
  Makhanek, A.G. and Skripko, G.A., 1973, 
  {\it Phys. Status Solidi (B)} {\bf 58}, 745.

\item{6.} Makhanek, A.G. and Skripko, G.A., 1979, 
{\it Phys. Status Solidi (A)}, {\bf 53}, 243. 

\item{7.} Jugurian, L.A., 1980, {\it Reports N$^\circ$ 232 and 
233}, Inst. Fiz. Akad. Nauk BSSR, Minsk. 

\item{8.}
  Makhanek, A.G., Korol'kov, V.S., and Yuguryan, L.A., 1988, 
  {\it Phys. Status Solidi (B)}, {\bf 149}, 231. 

\item{9.} Kibler, M. and G\^acon, J.-C., 1989, 
{\it Croat. Chem. Acta}, {\bf 62}, 783. 

  \item{10.} 
  Kibler, M.R., 1991, 
  {\it Symmetry and Structural Properties of Condensed Matter},
  Eds. Florek, W., Lulek, T., and Mucha, M. 
  (Singapore: World Scientific). 

  \item{11.} 
  Daoud, M.~and Kibler, M., 1992, {\it Laser Phys.}, {\bf 2}, 704. 

  \item{12.} 
  Daoud, M.~and Kibler, M., 1993, {\it J.~Alloys and 
  Compounds}, {\bf 193}, 219. 

  \item{13.} 
  Kibler, M. and Daoud, M., 1993, {\it Lett. Math. Phys.}, {\bf xx}, xxx. 

  \item{14.} 
  Sztucki, J., 1993, {\it Chem. Phys. Lett.}, {\bf 203}, 383. 

\item{15.} G\^acon, J.C., Marcerou, J.F., Bouazaoui, M., 
Jacquier, B., and Kibler, M., 1989, {\it Phys. Rev. B}, 
{\bf 40}, 2070. 

\item{16.} 
Gâcon, J.C., Jacquier, B., Marcerou, J.F., Bouazaoui, M., 
and Kibler, M., 1990, {\it J. Lumin.}, {\bf 45}, 162.

\item{17.} 
G\^acon, J.C., Bouazaoui, M., Jacquier, B., Kibler, M., 
Boatner, L.A., and Abraham, M.M., 1991, 
{\it Eur. J. Solid State Inorg. Chem.}, {\bf 28}, 113. 

\item{18.} 
Gâcon, J.C., Burdick, G.W., Moine, B., and Bill, H., 
1993, {\it Phys. Rev. B}, {\bf 47}, 11712.

\item{19.} 
Sztucki, J., Daoud, M., and Kibler, M., 1992, 
{\it Phys. Rev. B}, {\bf 45}, 2023. 

\item{20.} 
Daoud, M. and Kibler, M., 1992, 
{\it J. Alloys and Compounds}, {\bf 188}, 255. 

\item{21.} Kibler, M., 
1968, {\it J. Mol. Spectr.},          {\bf 26},  111~; 
1969, {\it Int. J. Quantum Chem.},    {\bf 3},  795~;
1969, {\it C.R. Acad. Sc. (Paris) B}, {\bf 268}, 1221.

\item{22.} Kibler, M.R., 1979, {\it Recent Advances in Group 
Theory and Their Application to Spectroscopy}, Ed. Donini, J.C. 
(New York: Plenum Press, New York). 

\item{23.} Daoud, M., 1992, {\it Doctorate Thesis}, Lyon-1 
University, France. 
 
\bye